\documentclass[%
 reprint,
superscriptaddress,
 amsmath,amssymb,
 aps,
pre,
]{revtex4-1}

\usepackage{graphicx}
\usepackage{dcolumn}  
\usepackage{bm}       
\usepackage{url}

\begin{document}

\title{Mapping and discrimination of networks in the complexity-entropy plane}

\author{Marc Wiedermann}
\email{marcwie@pik-potsdam.de}
\affiliation{Potsdam Institute for Climate Impact Research --- Telegraphenberg
  A31, 14473 Potsdam, Germany, EU}
\affiliation{Department of Physics, Humboldt University --- Newtonstr. 15,
	12489 Berlin, Germany, EU}
\author{Jonathan F. Donges}
\affiliation{Potsdam Institute for Climate Impact Research --- Telegraphenberg
  A31, 14473 Potsdam, Germany, EU}
\affiliation{Stockholm Resilience Centre, Stockholm University --- Kr\"aftriket 2B, 114 19 Stockholm, Sweden, EU} 
\author{J\"urgen Kurths}
\affiliation{Potsdam Institute for Climate Impact Research --- Telegraphenberg
  A31, 14473 Potsdam, Germany, EU}
\affiliation{Department of Physics, Humboldt University --- Newtonstr. 15,
	12489 Berlin, Germany, EU}
\author{Reik V. Donner}
\affiliation{Potsdam Institute for Climate Impact Research --- Telegraphenberg
  A31, 14473 Potsdam, Germany, EU}

\date{\today}

\begin{abstract}
  Complex networks are usually characterized in terms of their topological,
  spatial, or information-theoretic properties and combinations of the
  associated metrics are used to discriminate networks into different classes
  or categories. However, even with the present variety of characteristics at
  hand it still remains a subject of current research to appropriately quantify
  a network's \textit{complexity} and correspondingly discriminate between
  different types of complex networks, like infrastructure or social networks,
  on such a basis. Here, we explore the possibility to classify complex
  networks by means of a statistical complexity measure that has formerly been
  successfully applied to distinguish different types of chaotic and stochastic
  time series.  It is composed of a network's averaged per-node entropic
  measure characterizing the network's information content and the associated
  Jenson-Shannon divergence as a measure of disequilibrium. We study 29
  real world networks and show that networks of the same category tend to
  cluster in distinct areas of the resulting complexity-entropy plane. We
  demonstrate that within our framework, connectome networks exhibit among the
  highest complexity while, e.g, transportation and infrastructure networks
  display significantly lower values. Furthermore, we demonstrate the
  utility of our framework by applying it to families of random scale-free
  and Watts-Strogatz model networks. We then show in a second application that
  the proposed framework is useful to objectively construct threshold-based
  networks, such as functional climate networks or recurrence networks, by choosing the
  threshold such that the statistical network complexity is maximized.
\end{abstract}

\maketitle

\section{Introduction}
Many real world systems are well represented by complex
networks~\cite{boccaletti_complex_2006,newman_structure_2003}. Examples include
social systems, such as herds of one or more species of
animals~\cite{croft2008exploring,mcgregor2005animal}, transportation systems,
such as road networks~\cite{kaiser_spatial_2004,gastner_spatial_2006}, or
connectome networks, such as the human
brain~\cite{bullmore2009complex,sporns2005human}. 

The structure of such networks is usually quantified by a set of
topological~\cite{albert2002statistical}, spatial~\cite{barthelemy2011spatial}
or information-theoretic~\cite{dehmer2011history} characteristics which measure
certain properties of either distinct nodes (local characteristics) or the
entire network itself (global characteristics). Specifically the latter may be
used to compare different kinds of networks as well as for categorizing a given
set of networks into different classes~\cite{costa_characterization_2007}. The
most prototypical example of such a discrimination would be the assignment of
the small world property to a given network (following Watts and Strogatz)
depending on the numerical values of its clustering coefficient and average
path length~\cite{watts1998collective}.

Other approaches have successfully distinguished between different classes of
scale-free networks by means of characteristics associated with their degree
distribution~\cite{goh2002classification}, or spatial networks by determining
bias corrected versions of macroscopic network
characteristics~\cite{wiedermann2016spatial}. Further, networks have been
assigned to so-called superfamilies based on the distribution of certain motifs
that form their substructure~\cite{milo2004superfamilies}. One successful
approach to quantify topological differences in networks of different types is
based on examining their community structure and yields statistical properties
within the communities that are unique to different types of networks under
study~\cite{lancichinetti_characterizing_2010}. 

However, while the large variety of present metrics allows for a quantification
of a network's particular macroscopic and microscopic structure, it still remains a subject of current
research to (i) assess the actual \textit{complexity} of a network based on
these sets of characteristics~\cite{anand_entropy_2009}, and (ii) to determine
distinct sets of properties for certain classes of networks, such as
infrastructure or social networks, in order to objectively and comprehensively
distinguish between them.  While there exists a variety of such complexity
measures~\cite{bonchev2005quantitative}, most of them are tailored to specific
applications and have so far not been successfully applied to intercompare
different types or classes of networks as in this respect they often lack a meaningful
interpretation~\cite{dehmer2009large}.

Contributing to the above issues, we introduce here a two-dimensional metric
based on an entropic and an adjoint statistical complexity measure to
distinguish different types of complex
networks~\cite{rosso2007noise,martin2006generalized}. This approach was
originally introduced to distinguish chaotic from stochastic systems in time
series analysis and has been successfully applied to study, e.g., ordinal
patterns in daily stream time series of river runoff~\cite{lange_ordinal_2013}.
Its purpose is to assign each system under study a position in a two
dimensional space spanned by an entropy and a statistical complexity measure,
the latter being a product of entropy and Jenson-Shannon divergence with
respect to a uniform distribution.

Here, we transfer this concept from time series to the case of complex networks
and redefine the above entropy and statistical complexity accordingly. Various
definitions of network entropies or, more specifically, the underlying
probability distributions have already been proposed. They may for example be
computed in terms of the network's topological information
content~\cite{rashevsky1955life} or, quite commonly, its degree
distribution~\cite{bonchev1983information,cancho_optimization_2003,wang_entropy_2006}.
Further definitions of entropy are based on the assessment of network ensembles
or randomized correspondents
thereof~\cite{bianconi_entropy_2008,anand_entropy_2009}. However, particularly
entropy measures that are based on the degree-distribution alone have been
shown to have little discriminative power when applied to a heterogeneous set
of graphs~\cite{dehmer_information_2012}.  It is in contrast rather advisable
to rely on local per-node definitions of network
entropies~\cite{dehmer_information_2012,konstantinova_applications_2003}.

One candidate for such a node-wise definition of entropy is based on
the probability of a random walker to jump from a specific node to its neighbors in the
network~\cite{small2013complex}. This notion of entropy is closely related to
random walks which themselves are in their application and interpretation
closely related with the assessment of a networks' navigability and thus,
complexity~\cite{barthelemy2011spatial,amaral2004complex,white2002navigability}.

We apply our formalism to 29 real world networks that are discriminated by
context into the four types of social animal, social affiliation,
transportation and connectome networks. We show that for most cases the
different types occupy distinct areas in the complexity entropy-plane. Thus,
our formalism naturally distinguishes between different types of systems under
study. We also apply the framework to two generic classes of benchmark
networks, namely an ensemble of scale-free networks with varying power law
exponents and a set of networks constructed from the Watts-Strogatz model for different
rewiring probabilities~\citep{watts1998collective}.  In a second application,
we show that our complexity measure can be used to objectively construct
threshold-based networks such as functional climate
networks~\cite{tsonis2006networks,donges2009complex} or recurrence
networks~\cite{donner2010recurrence} by choosing a discrete network
representation that maximizes statistical complexity.

The remainder of this paper is organized as follows:
Section~\ref{sec:methods_data} presents the methodology that is put
forward in this work and additionally introduces the kinds of networks that
are studied in the two example applications. The corresponding results are presented in
Sec.~\ref{sec:results} and the work is ultimately concluded with an outlook in
Sec.~\ref{sec:outlook}.

\section{Methods \& Data}\label{sec:methods_data}
An unweighted network $G$ with a set of $N$ nodes labeled with integers
$i=1,\ldots, N$ and corresponding links between nodes can be represented by its
$N\times N$ adjacency matrix $\textbf A$ with entries $A_{ij} = 1$ if nodes $i$
and $j$ are connected by an edge, and $A_{ij} = 0$ otherwise. Each node $i$'s
number of directly connected neighbors $k_i$ is computed as $k_i=\sum_j A_{ij}$
and is referred to as the degree of node $i$. For this work, we further assume
undirected networks with no self-loops. Thus, $\textbf A = \textbf A^\text{T}$,
$A_{ii}=0\ \forall\ i=1,\ldots, N$, and $k_i \leq N-1$.

Analogously to the concept of a \textit{complexity-entropy plane}
in nonlinear time series analysis, which has been utilized to discriminate
between different types of time series generated by stochastic and deterministic
chaotic processes~\cite{rosso2007noise,martin2006generalized}, we aim to
characterize a set of complex networks by means of its average per-node Shannon
entropy $S$ and a statistical complexity measure $C$. We thereby make use of
two notions that are related with the complexity of a physical
system, namely its \textit{information content} and its state of
\textit{disequilibrium}~\cite{lopez-ruiz_statistical_1995,calbet2001tendency,lamberti_intensive_2004}.
In particular, we relate the information content of the network with the
entropy $S$ and the disequilibrium with the network's Jenson-Shannon divergence
$Q$ with respect to an appropriately chosen reference state. 

Before going to the case of complex networks we discuss as a preliminary and as
an analogy to classical statistical physics the two most extreme cases of
complexity one might consider, namely the crystal as well as the ideal gas
displaying a large and a low degree of order,
respectively~\cite{anderson1991complexity,parisi1993statistical,lopez-ruiz_statistical_1995}.

It is easily deductible that due to its regular structure, the crystal usually
contains low or almost zero information, and, hence $S\rightarrow 0$. In
contrast to this, the ideal gas (due to its disorder) contains a large amount of
information, implying $S\gg0$.  Further, it is observed, that the perfect
crystal displays among the highest disequilibrium ($Q\gg0$), i.e., a large
degree of order, while the ideal gas displays the exact opposite ($Q\rightarrow
0$). As both measures, $S$ and $Q$, usually increase or decrease monotonically
with a system's complexity, we ultimately derive a measure of statistical
complexity $C$ as the product of both, information content (e.g, Shannon
entropy $S$) and disequilibrium (e.g, Jenson-Shannon divergence
$Q$)~\cite{lopez-ruiz_statistical_1995}. In the following we
transfer the notion of information content and disequilibrium to the case of
complex networks and derive corresponding terms for the entropy $S$ and the
statistical complexity $C$ that then ultimately form the two-dimensional
complexity-entropy plane. 

\subsection{Network entropy}\label{sec:entropy}
Generally, the classical Shannon entropy for discrete probability distributions
is given by
\begin{align}
  S(P) = -\sum_k p_k \log p_k.
\end{align}
Here, $p_k$ denotes the probability of occurrence for a given state $k$.  Since
averaged per-node entropies have been shown to generally serve as a good choice
for discriminating between different types of networks~\cite{dehmer_information_2012},
we choose here one specific definition of entropy that is based on the
assessment of probabilities to jump between nodes when randomly traveling
through the network and that has been successfully applied to the study of
complex networks constructed from univariate time
series~\cite{small2013complex}. In particular, the entropy $S_i$ for each node $i$ is computed
based on the distribution $P_i$ with entries $p_{i\rightarrow j}$ that give the
uniformly distributed probability to jump from node $i$ to node $j$ along an edge between them in
exactly one step. Thus, the corresponding random walk is formulated analogously to
its application in computing the recently proposed $\textit{random walk
  betweenness}$~\cite{newman2005measure}. If a node $i$ is not fully
disconnected from the rest of the network (i.e., $k_i>0$), the corresponding probabilities
$p_{i\rightarrow j}$
are given by
\begin{align}
  p_{i\rightarrow j} = \frac{A_{ij}}{k_i} \in \{0, 1/k_i\} \label{eqn:p_ij}
\end{align}
with $\sum_j p_{i\rightarrow j} = 1$.
The node entropy then reads,
\begin{align}
  S_i(P_i) &= -\sum_{j=1}^N p_{i\rightarrow j} \log p_{i \rightarrow
    j}\nonumber\\
  &= -\sum_{j} \frac{A_{ij}}{k_i} \log \frac{A_{ij}}{k_i} = \log k_i.
\end{align}
In case of an isolated node $i$ with $k_i=0$, we set $S_i(P_i)=0$.
Ultimately, the average normalized entropy taken over all nodes $i$ is
referred to as the network entropy,
\begin{align}
  S(P) = \frac{1}{N\log (N-1)} \sum_{i} S_i(P_i) \in [0, 1].\label{eqn:entropy}
\end{align}
This specific definition of entropy is in accordance with some magnitude-based
information indices~\cite{bonchev1983information} that measure tendencies for
complex networks to form branches. In particular, our measure quantifies the
heterogeneity in the network's degree distribution in a sense that nodes with
low degree, i.e., peripheral nodes lower the overall network entropy $S(P)$
while high degree-hubs cause its increase. Thus, the present definition of $S(P)$
incorporates not only average statistics of the network's degree distribution
but implicitly also accounts for its higher moments, such as the variance. The
entropy $S(P)$ can further be interpreted with respect to the underlying
formulation of the random walk. 

In the limiting case of a fully connected
network the probability to jump between nodes is given as $p_{i\rightarrow j} =
\frac{1}{N-1} \ \forall\ i\neq j$. Thus, the walk becomes fully random in a
sense that no node $j$ is excluded as a possible candidate for the walker to jump
to. This case is directly related to the notion of the ideal gas 
outlined above where all micro states are equally probable and, thus, the
entropy is maximized. Analogously, for the fully connected network all walks
through the network of arbitrary length are equally probable, too.
Consequently, the entropy $S(P)$ is also maximized and reads $S(P)=1$. 

In turn, for a sparsely connected network the jumps of the walker become more
deterministic and in the limiting case of, e.g., node $i$ only having one
neighbor $n$, its associated traverse probabilities approach $p_{i \rightarrow
  j} = \delta_{jn}$ (with $\delta_{jn}$ being Kronecker's delta).  In this case,
the walker has only one option for jumping to a neighboring node of $i$.
Consequently, the entropy is lowered and for sufficiently sparse networks
approaches $S(P)\rightarrow 0$. Again, this case may be interpreted in analogy
to a regular crystal that displays perfect order as well as a deterministic
structure and, hence, has a low information content.

In summary, we thus interpret our definition of  $S(P)$ as a measure of
regularity or order in the network under study with respect to its navigability
that is measured in terms of a random walk.  

\subsection{Statistical complexity}
We aim to express statistical complexity or non-triviality in terms of a
system's disequilibrium and information
content~\cite{lopez-ruiz_statistical_1995}, with the latter being defined as
the network entropy $S(P)$ that is introduced above. Along the lines of common statistical
mechanics, disequilibrium is conveniently measured in terms of the Jenson-Shannon divergence~\cite{martin2003statistical}
\begin{align}
  Q_i(P_i, P_{i, e}) = Q_0 &\{ S_i(0.5[P_i + P_{e, i}]) \nonumber\\&-
  0.5[S_i(P_i) + S_i(P_{e, i})]\},
\end{align}
with $Q_0=1 / \log 2$ to ensure $Q_i\in [0,1]$. This metric takes low values
for systems that are close to equilibrium like an ideal gas and high values
for systems in disequilibrium like the perfect crystal. Here, the
probability distribution $P_i$ with entries as introduced in
Eq.~\eqref{eqn:p_ij} again denotes the probabilities to jump between
neighboring nodes $i$ and $j$ when randomly traveling through the network. The
distribution $P_{e, i}$ denotes the same, but for an appropriately chosen
reference or equilibrium state, i.e., network. Analogously to previous works, we
assume that a system is in equilibrium if its state corresponds to the fully
randomized one~\cite{martin2003statistical}. For the case of complex networks,
this equilibrium state would than be a corresponding Erd\H
os-R\'enyi~\cite{erdos_evolution_1960} random network. We introduce the
specific details of this choice in Sec.~\ref{sec:reference_networks}.

Analogously to the network entropy, the Jenson-Shannon divergence $Q$ of the
entire network is again computed as the arithmetic mean of all per-node values
$Q_i$,
\begin{align}
  Q(P, P_e) = \frac{1}{N} \sum_i Q_i(P_i, P_{i, e}).
\end{align}

As for the case of the network entropy $S(P)$, the analogy with
generic physical systems (perfect crystal and ideal gas) is apparent. 

Again, the fully connected network corresponds to the case of an ideal gas with
minimum disequilibrium as any appropriately chosen reference network should be
fully connected as well, which implies $P=P_e$ and, thus, $Q = 0$. In contrast,
a randomly chosen reference to a sparsely connected network most certainly
displays a different microscopic structure. Hence, the probabilities $P$ and $P_e$ for
jumping between nodes also differ, yielding a high disequilibrium
$Q \gg 0$.

With the above observations in mind, we demand based on common sense that neither the fully connected
nor the very sparsely connected (or almost empty) network should be attributed a
large complexity. Hence, neither a measure of $\textit{information}$ ($S(P)$)
nor $\textit{disequilibrium}$ $Q(P, P_e)$ alone may serve as an appropriate
quantifier of statistical complexity. However, a measure
of statistical complexity $C$ has been proposed that is based on a product of the two
quantities~\cite{lopez-ruiz_statistical_1995,rosso2007noise},
\begin{align}
  C(P) = Q(P, P_e) S(P) \in [0, 1].
\end{align}
This measure intuitively exhibits the required asymptotic properties, such that for the
limiting case $S(P)=0$, it follows that $C(P)=0$.  Analogously, $S(P)=1$ is only
achieved for a fully connected network which implies $P=P_e$ (see
Sec.~\ref{sec:reference_networks} for details) and $Q(P, P_e)=0$, which also yields $C(P)=0$. For all cases
$0<S(P)<1$, the statistical complexity $C(P)$ has a possible upper bound that
is determined by $S(P)$. However, its analytical expression 
has so far only been obtained for a binary state probability
distribution~\cite{lamberti_intensive_2004}. 

We ultimately note that a variety of further measures has been developed that
similarly aim to quantify complexity in dynamical
systems~\cite{wackerbauer1994comparative}. However, most of these measures are
more tailored to other applications, such as the numerical 
detection of bifurcations, e.g, order-chaos or chaos-chaos transitions. We thus focus in this work on the statistical
complexity measure as introduced above.

\subsection{Reference networks}\label{sec:reference_networks}
\begin{figure}[t!]
  \centering \includegraphics[width=.9\linewidth]{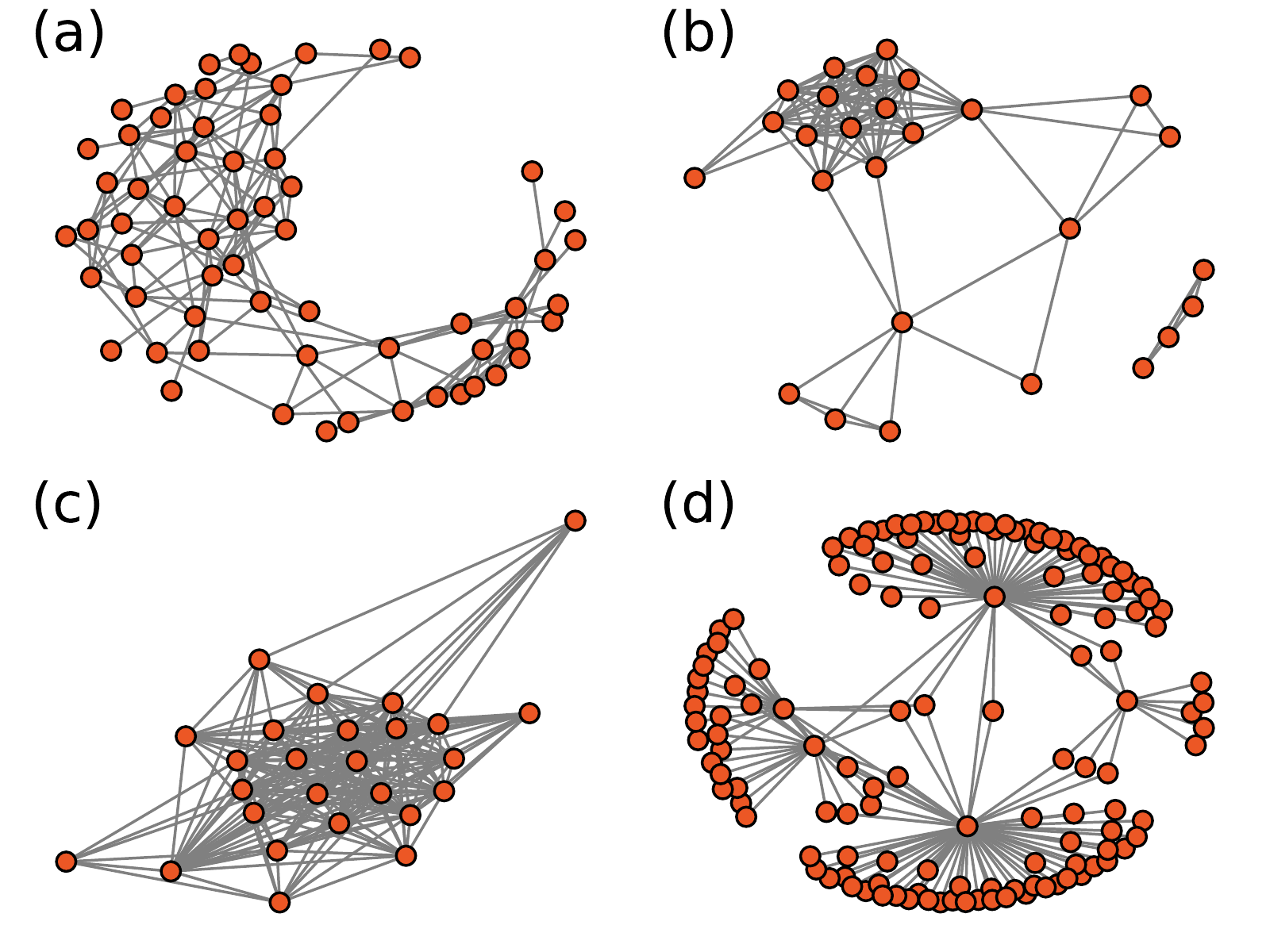}
  \caption{Selection of four out of the 29 networks investigated in this study:
    Dolphin (a), zebra (b) and bison (c) animal social networks as well as the
    network of American revolutionary groups (d).}\label{fig:networks}
\end{figure}

In order compute the Jenson-Shannon divergences $Q_i$ and $Q$  we need to
compare each network's set of probability distributions $P_i$ to jump between a node
$i$ and its neighbors with an
appropriately chosen baseline or equilibrium $P_{i, e}$. We note here that
defining the equilibrium state of a network is a highly non-trivial task that
is often achieved by fitting the network under study to a certain network model
using numerical variational techniques in order to minimize or maximize a
target or cost
function~\cite{dafermos_sensitivity_1984,giannessi_variational_1995}. However,
in order to demonstrate the applicability of our approach and to focus on the
numerical properties of the statistical complexity $C$, we chose to define the
equilibrium or reference state of a given network as its fully randomized
counterpart and, thus, interpret $C$ as an indicator of statistical independence
from a corresponding random state. 
In this case, one obvious candidate for such a baseline network is
the Erd\H os-R\'enyi random graph~\cite{erdos_evolution_1960}. To account for
the stochasticity of this model, we generate for each network under study an
ensemble of $n=100$ independent Erd\H os-R\'enyi networks with the same number
of nodes $N$ and linking probability $\rho=\sum_i k_i / (N(N-1))$ and compute
ensemble average quantities of $Q_i$ and $Q$ from the resulting probability
distributions $P_{i, e}$. 

\subsection{Real-world networks}
\begin{table*}[t]
\centering
\setlength{\tabcolsep}{.2cm}
\begin{tabular}{llrccc}
Name & Category & $N$ & $\rho$ & $S$ & $C$ \\\hline
Sheep \cite{hass_social_1991} & Social Animal & 28 & 0.622 & 0.825 & 0.326 $\pm$ 0.022\\
Rhesus \cite{sade_sociometrics_2004} & Social Animal & 16 & 0.575 & 0.753 & 0.341 $\pm$ 0.035\\
Kangaroo \cite{grant_dominance_1973} & Social Animal & 17 & 0.669 & 0.782 & 0.285 $\pm$ 0.024\\
Mac \cite{fedigan_monkeys_1991} & Social Animal & 62 & 0.617 & 0.874 & 0.339 $\pm$ 0.009\\
Bison \cite{lott_dominance_1979} & Social Animal & 26 & 0.683 & 0.855 & 0.281 $\pm$ 0.019\\
Zebra \cite{sundaresan_network_2007} & Social Animal & 27 & 0.316 & 0.571 & 0.409 $\pm$ 0.024\\
Cattle \cite{schein_social_1955} & Social Animal & 28 & 0.542 & 0.772 & 0.372 $\pm$ 0.018\\
Dolphins \cite{lusseau_bottlenose_2003} & Social Animal & 62 & 0.084 & 0.342 & 0.317 $\pm$ 0.006\\
Autobahn \cite{kaiser_spatial_2004} & Transportation & 1,168 & 0.002 & 0.099 & 0.099 $\pm$ 0.000\\
USairport500 \cite{colizza_reactiondiffusion_2007} & Transportation & 500 & 0.024 & 0.250 & 0.246 $\pm$ 0.001\\
USairport 2010 \cite{noauthor_why_2011} & Transportation & 1,574 & 0.014 & 0.248 & 0.246 $\pm$ 0.000\\
Openflights \cite{noauthor_why_2011} & Transportation & 2,939 & 0.004 & 0.173 & 0.173 $\pm$ 0.000\\
Rome99 \cite{noauthor_9th_nodate} & Transportation & 3,353 & 0.001 & 0.121 & 0.121 $\pm$ 0.000\\
South-Africa \cite{hobson_evolution_1906} & Social Affiliation & 6 & 0.633 & 0.601 & 0.306 $\pm$ 0.062\\
American Revolution \cite{noauthor_american_nodate} & Social Affiliation & 136 & 0.017 & 0.043 & 0.043 $\pm$ 0.001\\
Club-Membership \cite{faust_centrality_1997} & Social Affiliation & 25 & 0.305 & 0.576 & 0.415 $\pm$ 0.021\\
Corporate-Leadership \cite{barnes_structural_2010} & Social Affiliation & 24 & 0.322 & 0.590 & 0.417 $\pm$ 0.026\\
Rhesus Brain 1 \cite{harriger_rich_2012} & Connectome & 242 & 0.105 & 0.523 & 0.474 $\pm$ 0.002\\
Rhesus Brain 2 \cite{markov_anatomy_2014} & Connectome & 91 & 0.142 & 0.452 & 0.401 $\pm$ 0.006\\
Mouse Retina 1 \cite{helmstaedter_connectomic_2013} & Connectome & 1,076 & 0.157 & 0.693 & 0.594 $\pm$ 0.001\\
Mixed Species Brain 1 \cite{reus_rich_2013} & Connectome & 65 & 0.351 & 0.717 & 0.478 $\pm$ 0.012\\
Rhesus Cerebral Cortex 1 \cite{markov_weighted_2014} & Connectome & 91 & 0.342 & 0.710 & 0.485 $\pm$ 0.007\\
C Elegans Neural Male 1 \cite{jarrell_connectome_2012} & Connectome & 269 & 0.081 & 0.486 & 0.451 $\pm$ 0.002\\
Rattus Norvegicus Brain 3 \cite{bota_online_2007} & Connectome & 493 & 0.214 & 0.684 & 0.558 $\pm$ 0.001\\
Rhesus Interareal Cortical Network 2 \cite{markov_role_2013} & Connectome & 93 & 0.529 & 0.822 & 0.413 $\pm$ 0.006\\
Rattus Norvegicus Brain 2 \cite{bota_online_2007} & Connectome & 502 & 0.196 & 0.666 & 0.556 $\pm$ 0.001\\
Rattus Norvegicus Brain 1 \cite{bota_online_2007} & Connectome & 503 & 0.182 & 0.653 & 0.553 $\pm$ 0.001\\
Mouse Brain 1 \cite{oh_mesoscale_2014} & Connectome & 213 & 0.716 & 0.934 & 0.272 $\pm$ 0.003\\
C Elegans Herm Pharynx 1 \cite{varshney_structural_2011} & Connectome & 279 & 0.059 & 0.460 & 0.436 $\pm$ 0.002\\
\hline
\end{tabular}
\caption{Overview of the networks evaluated in this study together with their
  respective number of nodes $N$ and link density $\rho$ as well as entropy
  $S$ and statistical complexity $C$ computed over an ensemble of $n=100$ reference
  networks. The provided estimate of the error in $C$ denotes one standard
  deviation. Categories have been
  assigned according to their classification in the Colorado Index of
  Complex Networks (ICON) (\url{https://icon.colorado.edu/}).}
\label{tab:tab_01}
\end{table*}

We study the entropy $S$ and statistical complexity $C$ of 29 real
world networks, for which we assign types according to their
sub-domains in the Colorado Index of Complex Networks (ICON)
(\url{https://icon.colorado.edu/}). Specifically, we study eight networks that
represent social networks among different species of animals, five
transportation networks, four networks representing affiliations between people
or corporations, and twelve connectome networks for different species. In order
to make the results comparable, we
treat all networks under study as being unweighted, undirected, and without
self-loops. The networks
under study, together with their assigned type, number of nodes $N$, and link
density $\rho$ are summarized in Tab.~\ref{tab:tab_01}. Visual
representations of the topological structure of four of the 29 networks are
shown in Fig.~\ref{fig:networks}. The network parameters $N$ and
$\rho$ will be used to compute corresponding reference networks as outlined in
Sec.~\ref{sec:reference_networks}.

\subsection{Threshold-based networks}\label{sec:functional_networks}
In addition to real world networks, we aim to illustrate the
usefulness of the statistical complexity $C$ as a measure to objectively
construct threshold-based networks. Generally, these types of networks are
constructed from $N\times N$ matrices that describe some spatial or similarity
relationship between nodes~\cite{achard_efficiency_2007}. 

We first study one prototypical example of a threshold-based network in
terms of a recurrence network~\cite{donner2010recurrence,lange_recurrence_2015}
constructed from the three-dimensional R\"ossler system given by
\begin{align}
  \frac{dx}{dt} &= -y -z\label{eqn:roesslerx}\\
  \frac{dy}{dt} &= x + ay\\
  \frac{dz}{dt} &= b + z(x-c).\label{eqn:roesslerz}
\end{align}
We set $a=b=0.2$ and $c=5.7$ as in the original study of this
system~\cite{rossler_equation_1976}. In the past, recurrence networks have been show to
capture essential information on the phase space structure of the dynamical system under
study and thus serve as a good (or even equivalent) representation of the
system's trajectory~\cite{donges2012analytical,donner2011recurrence}.
Each node $i$ in the network represents a
point $\vec x_i = (x(t_i), y(t_i), z(t_i))$ on the system's trajectory at
randomly chosen times $t_i\in [100, 1000]$, where $t_i\geq 100$ ensures that
for our choice of initial values $x(0)=y(0)=z(0)=1$ the system has converged
onto the chaotic attractor. The entries $D_{ij}$ of the distance matrix
$\textbf D$ are then given by the Euclidean distances between points $\vec x_i$
and $\vec x_j$~\cite{donner2010recurrence}. From $\textbf D$, a corresponding
recurrence matrix $\textbf R$ with entries $R_{ij}$ is constructed by choosing
a recurrence threshold $T$ such that 
\begin{align}
  R_{ij} = \Theta(T - D_{ij}) \label{eqn:recurrence}
\end{align}
where $\Theta(\cdot)$ denotes the Heaviside function. $\textbf R$ is now
interpreted as the adjacency matrix of a spatial recurrence network such that
$A_{ij} = R_{ij} - \delta_{ij}$.  Hence, only distances between points that are
smaller than a critical distance $T$ are connected in the resulting network.
The threshold $T$ is chosen such that a desired link density or
\textit{recurrence rate} $\rho$ is obtained.

Another case of threshold-based networks are functional networks. Here, a
similarity matrix $\textbf M$ is constructed from pairwise statistical
interdependencies between time series that are represented by nodes
in the network. These nodes may correspond to different channels of
electroencephalography (EEG) signals in
neural networks~\cite{bullmore2009complex} or records of climatic variables
at different locations of the Earth in so-called climate
networks~\cite{donges2009complex,tsonis2006networks}. Specifically, the latter
have been shown to encode valuable information on the large-scale dynamical organization
of spatially extended components of the climate system, such as ocean
currents~\cite{donges_backbone_2009} 
or the El Ni\~no Southern
Oscillation~\cite{wiedermann2016climate}. As an example for such 
functional climate networks, we compute the pairwise Pearson correlation
between all $N=10,224$ time series of (i) monthly averaged surface air
temperature and (ii) monthly averaged sea level pressure from the NCEP/NCAR
40-year reanalysis project~\cite{kalnay1996ncep} that is provided by the
National Center of Oceanic and Atmospheric Research. Analogously to
Eq.~\eqref{eqn:recurrence}, a threshold is applied to the thus obtained
similarity matrix $\textbf M$  (containing the absolute values of the pairwise
Pearson correlations, i.e., $M_{ij}=|C_{ij}|$), such that only a
certain fraction of the largest values are considered as links in the resulting
network. Therefore,
\begin{align}
  A_{ij} = \Theta(M_{ij} - T) \cdot (1-\delta_{ij}),\label{eqn:climate}
\end{align}
with $A_{ij}$ being the entries of the resulting adjacency matrix $\textbf A$.
Again, the threshold $T$ is usually chosen such that a desired network link
density $\rho$ is achieved.

\section{Results}\label{sec:results}
\begin{figure} \centering
  \includegraphics[width=0.85\linewidth]{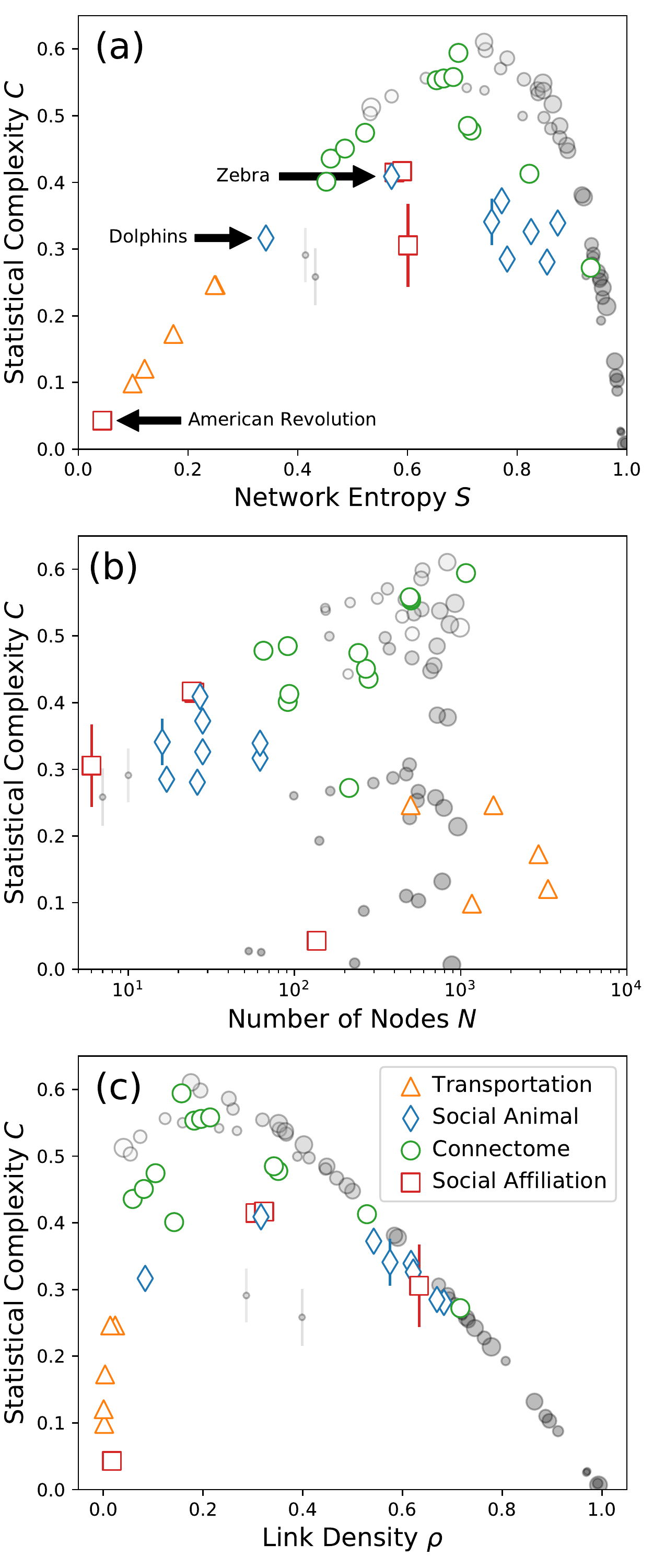} \caption{(a)
    Mapping of real world networks in the complexity-entropy
    plane. Additionally, gray scatter show the results for 50
    different Erd\H os-R\'enyi random networks. Here, the
    size and transparency denotes the uniformly at random drawn number of nodes
    $N$ from the interval $[10, 1000]$
    and linking probability $\rho$ from the interval $(0, 1]$, respectively. (b) Dependence of the statistical
    complexity $C$ on the number of nodes $N$ in each network under study. (c)
    The same as in (b) for the link density $\rho$. Error bars indicate one
    standard deviation of statistical complexity taken over the ensemble of
    $n=100$ random Erd\H os-R\'enyi reference networks with respect to the
    corresponding real world network under study and are shown if their size exceeds
    that of the corresponding symbol.}\label{fig:figure1}
\end{figure}

We now study in a first application the numerical values of entropy $S$ and
complexity $C$ for the different real world networks. To further consolidate
our findings we then also study two different classes of synthetic networks,
namely Watts-Strogatz networks with different rewiring
probabilities~\citep{watts1998collective} and random scale-free networks with a
prescribed exponent of the power law degree distribution. Ultimately, in a last use case, we illustrate
the application of statistical complexity to objectively determine appropriate
thresholds for the construction of threshold-based networks.

\subsection{Real world networks}
Figure~\ref{fig:figure1}(a) displays the entropy $S$ and average statistical complexity
$C$ of all real world networks under study with respect to ensembles of $n=100$
Erd\H os-R\'enyi reference networks reflecting the original networks'
respective properties (the average numerical values of $S$ and
$C$ are also presented in Tab.~\ref{tab:tab_01}). In addition, error bars
indicate the corresponding standard deviation taken over all ensemble members
and are shown when their size exceeds that of the corresponding symbol. For
reference, we also compute and
display the complexity and entropy of a set of 50 Erd\H os-R\'enyi networks
with the number of nodes $N$ and linking probability $\rho$ drawn uniformly at
random from the intervals $[10, 1000]$ and $(0, 1]$, respectively. 

We note that the different types of networks under study
generally occupy distinct areas in the complexity-entropy plane
(Fig.~\ref{fig:figure1}(a)). While connectome networks show among the
highest values of $C$, we note intermediate values for both types of social
networks, and the lowest values for the transportation networks. Additionally, the
latter also exhibit among the lowest values of entropy $S$. 
Notable exceptions are the social networks of dolphins and zebras, which in
contrast to most of the other animal networks display a unique community
structure (see Fig.~\ref{fig:networks}(a,b) for a visual representation).
Specifically, the dolphin network (Fig.~\ref{fig:networks}(a)) is characterized
by two distinct communities that are connected only via few nodes while the
zebra network (Fig.~\ref{fig:networks}(b)) is composed of one large almost
fully connected community containing roughly half of the nodes and at least two
further distinct communities with only few nodes that hardly connect with the
main herd. In contrast, all the other animal networks (see
Fig.~\ref{fig:networks}(c) for a representative example) generally display a
similar structure with only one densely connected community. Another
outstanding exception is the network of American revolutionary groups
(Fig.~\ref{fig:networks}(d)), which due to its distinct hierarchical community
structure displays very low values of entropy and complexity. We conclude from
these first observations that the complexity-entropy plane generally
distinguishes well between different types of networks solely based on their
specific and distinct topology. 

For the random Erd\H os-R\'enyi networks (gray symbols in
Fig.~\ref{fig:figure1}) we find that in many cases they show a higher statistical
complexity than real world networks. In fact, their values roughly seem to
determine an upper bound of $C$ for each possible value of $S$
(Fig.~\ref{fig:figure1}(a)).  This behavior is expected, as the two networks
that are compared in the Jenson-Shannon divergence $Q(P, P_e)$, the Erd\H
os-R\'enyi network under study and a random reference network, are
statistically fully independent by construction and, more importantly,
therefore less or equally less statistically dependent than any real world
complex network in comparison with a random reference network.  However, we
note that this property only seems to hold for sufficiently large networks
(Fig.~\ref{fig:figure1}(a)). 

Since the topological characteristics of the Erd\H os-R\'enyi network only
depend on the given number of nodes $N$ and
linking probability or link density $\rho$, we examine the dependence of $C$ on
both parameters individually. Figure~\ref{fig:figure1}(b) shows the values of
statistical complexity $C$ as a function of the number of nodes $N$ in each
network which displays no clear dependence between the two variables. In
contrast, a possible dependence between link density $\rho$ and statistical
complexity $C$ is observed (Fig.~\ref{fig:figure1}(c)). Still, we note that
networks with highly dissimilar link densities $\rho$ may exhibit similar
statistical complexity (Fig.~\ref{fig:figure1}(c)). 

Furthermore, the quantitatively similar functional dependencies between $S$ and $C$
(Fig.~\ref{fig:figure1}(a)) as well as $\rho$ and $C$ imply an expected
functional dependence between $S$ and $\rho$. However, the
$S$-$C$ plane is a much better choice for categorizing networks than the
$\rho$-$C$ plane since the entropy $S$ captures all moments in the degree
distribution of a given network (as can be seen from the series expansion of $\sum_i
\log k_i$ in Eq.~\eqref{eqn:entropy}), while $\rho$ only captures its first
moment.

\subsection{Synthetic networks}
\begin{figure} \centering
  \includegraphics[width=0.85\linewidth]{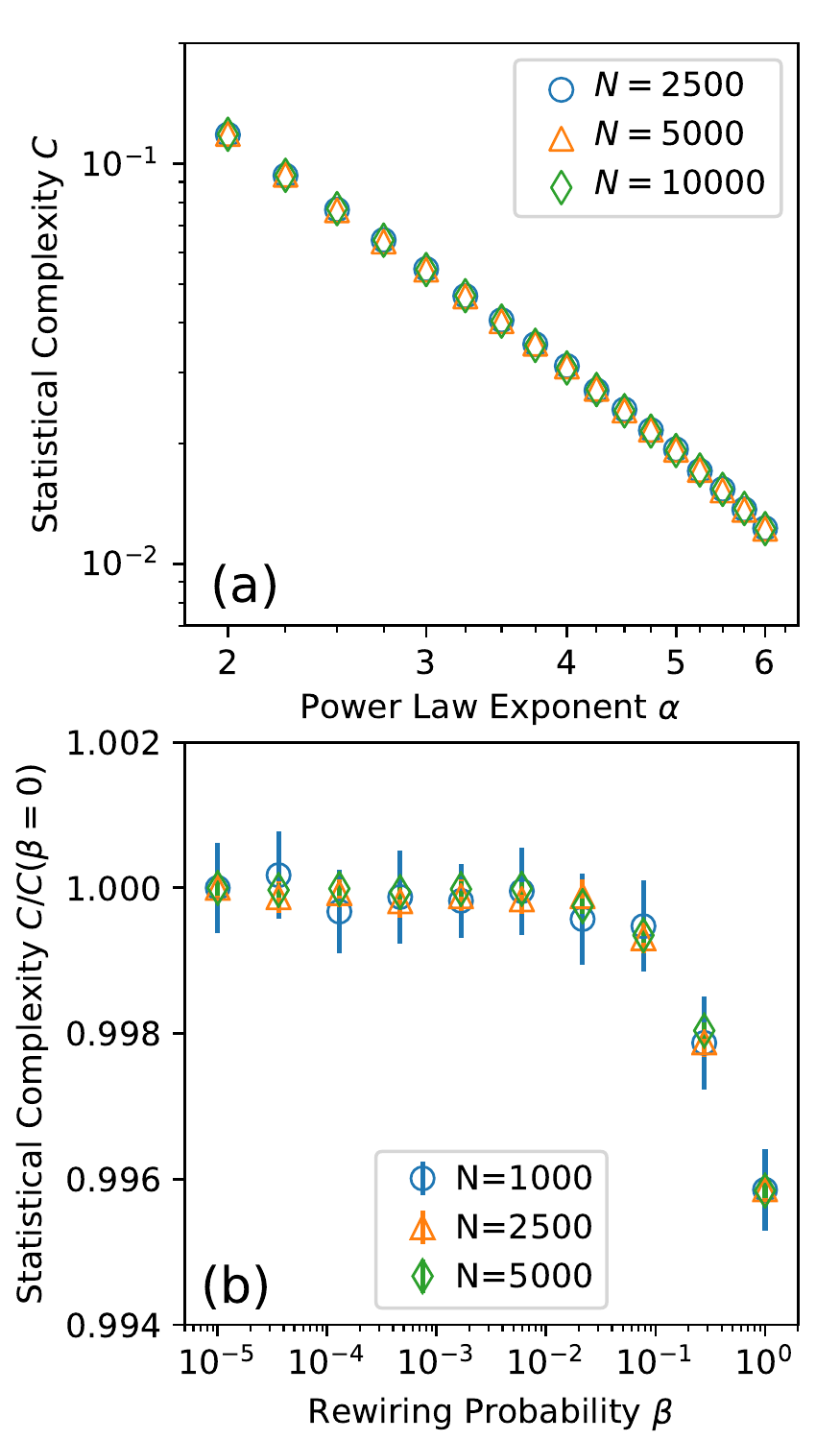} 
  \caption{(a) Average statistical complexity $C$ for different
    power law exponents $\alpha$ in the degree distribution of randomly
    generated scale-free networks. Each scatter denotes the average over an
    ensemble of 50 networks consisting of $N=2,500$, $N=5,000$ and $N=10,000$
    nodes, respectively. (b) The same for different choices of the rewiring
    probability $\beta$ in the Watts-Strogatz model with an average degree of
    $K=20$. For comparison all
    average statistical complexities are rescaled by 
    $C(\beta=0)$ of the regular ring graph. Error bars
    denote one standard deviation in the respective statistical complexity and are shown
    if their size exceeds that of the corresponding symbol.
  }\label{fig:pre_synthetic}
\end{figure}
To further consolidate the above findings we now systematically study
numerically
the statistical complexity $C$ for two different types of synthetic
networks, i.e., random scale-free and Watts-Strogatz networks.
Figure \ref{fig:pre_synthetic}(a) shows the statistical complexity $C$
averaged over an ensemble of $n=50$ random scale-free networks with power
law shaped degree distributions at different exponents $\alpha$ and different
numbers of nodes $N=2,500$, $N=5,000$ and $N=10,000$. In particular, each network
is constructed by first generating a degree sequence with a power law shaped
distribution according to the considered power law exponent $\alpha$. Then, the
network is created by iteratively inserting links between nodes according to
the configuration model~\citep{newman_structure_2003,boccaletti_complex_2006}.
Ultimately, self-loops and multiple links are again excluded from the assessment. 
We observe a decrease in statistical complexity $C$ with increasing $\alpha$.
For small values of $\alpha$ the networks display a heterogeneous degree
distribution with the presence of both, hubs and peripheral nodes.
Consequently, the statistical complexity $C$ takes comparatively large values.
In contrast, for high values of $\alpha$ the networks become increasingly
sparse with only few links present per node and, thus, they display a narrower degree distribution.
In this case the distinction between hubs and peripheral nodes is
less apparent and the network itself may be considered less statistically complex
which manifests in comparably low values of $C$. We also note, that due to the
scale-free property of the considered networks the statistical complexity $C$
seems to be
independent of the number of nodes $N$.

As a second family of model networks we study the statistical complexity $C$ for an ensemble of
networks constructed from the Watts-Strogatz model for different choices of the rewiring probability
$\beta$. Again, we construct networks of different sizes with $N=1,000$,
$N=2,500$ and $N=5,000$ nodes and a fixed average degree of $K=20$. Starting from
a ring graph where every node has each $K/2$ left and right neighbors, each link in
the network is rewired with probability $\beta$. As above, we obtain for each
choice of $N$ and $\beta$ an ensemble of $n=50$ randomly generated networks and
compute the corresponding average statistical complexity $C$
(Fig.~\ref{fig:pre_synthetic}(b)). In order to render the results comparable, we
rescale all obtained values by the corresponding statistical complexity of
the ring graph with $\beta=0$. The corresponding values of this
rescaling-factor read $C(\beta=0)=0.425$ for $N=1,000$, $C(\beta=0)=0.380$ for
$N=2,500$ and $C(\beta=0)=0.350$ for $N=5,000$, respectively. Thus, $C$ decreases with
increasing $N$ as the networks become more sparse. In contrast to the above
case of scale-free networks we note only minor, yet systematic, changes of $C$
with varying $\beta$ (Fig.~\ref{fig:pre_synthetic}(b)). In particular, the
observed drop in statistical complexity occurs for values of the rewiring
probability between $\beta=0.01$  and $\beta=0.1$. These values coincide with the
onset of the transition between small-world and random network structure in the
Watts-Strogatz model~\citep{watts1998collective}.

The observed small variations in $C$ may be 
explained from the underlying definition of the statistical complexity as a
result of a random walk between nodes in the network.  Rewiring the network
structure of a ring graph only induces minor changes to its degree
distribution, i.e., from a single peak at the average degree $K$ for $\beta=0$
to approximately a Poisson distribution centered around $K$ for $\beta=1$.
Since the random walk is to a large extent determined by the functional form of
this underlying distribution, resulting values of $C$ consequently vary only
little with $\beta$.  However, since we choose networks generated from the Erd\H
os-R\'enyi model as a reference, it is to be expected that a ring-graph 
displays a comparatively
larger complexity than its fully randomized counterpart ($\beta=0$) which
corresponds to a network obtained from the Erd\H os-R\'enyi model itself.
Ultimately, we note that by rescaling the obtained values of $C$ with
corresponding values of the ring graph, the relative changes in $C$ with
varying $\beta$ are largely similar for all choices of $N$. 

\subsection{Threshold-based networks}
\begin{figure} \centering
  \includegraphics[width=0.99\linewidth]{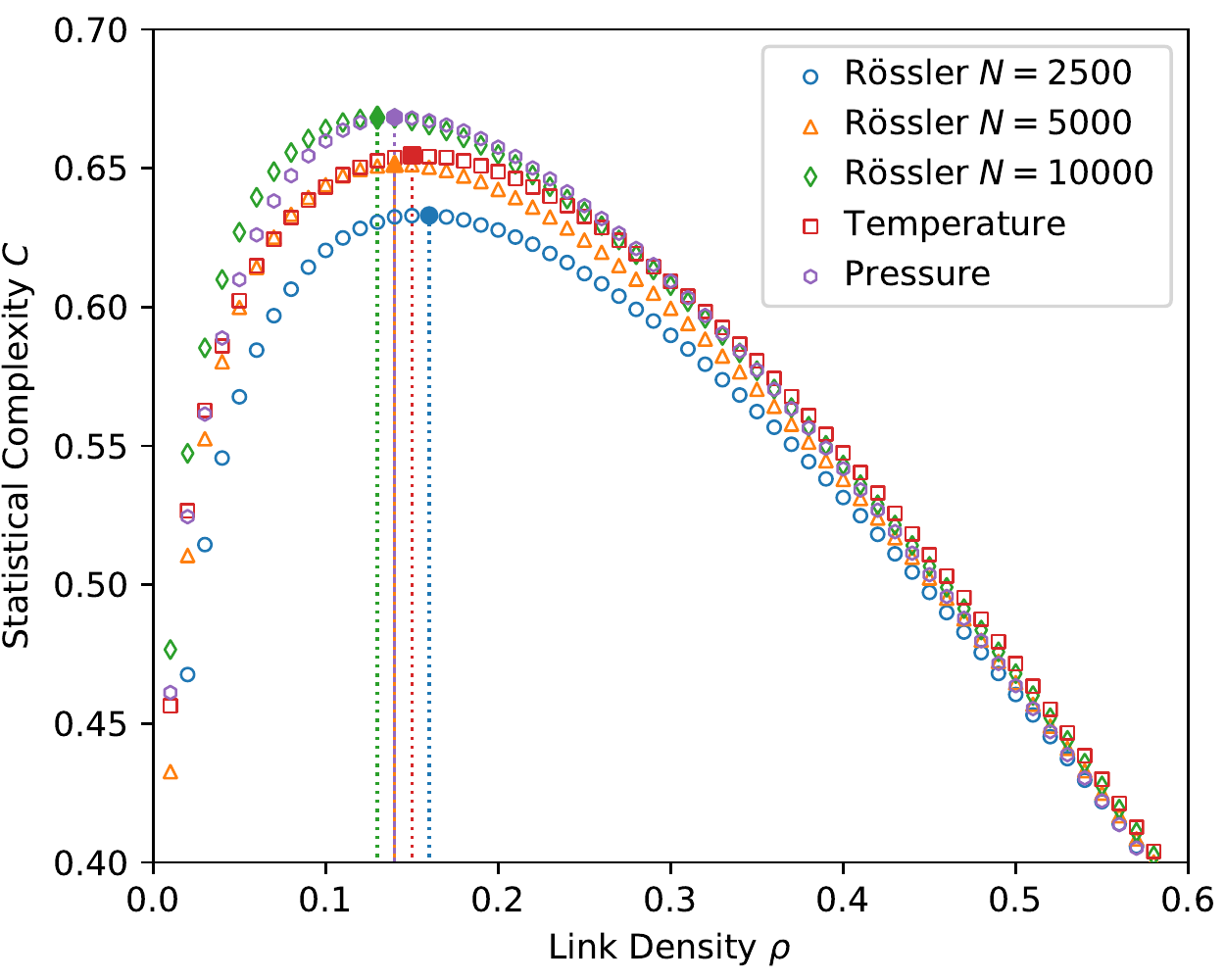} 
  \caption{Statistical complexity $C$ depending on the threshold-based networks'
    link densities $\rho$ for three recurrence networks with different numbers
    of nodes $N$ constructed from the R\"ossler system and two functional
    climate networks representing surface air temperature and sea level
    pressure variations, respectively. Filled symbols denote the maximum value of
    statistical complexity for each network. Dotted lines indicate the
    corresponding link density $\rho_{max}$, which maximizes the
    statistical complexity. No error bars are shown as the standard deviation
    of $C$ taken over all $n=100$ reference networks is always smaller than the size of
    the symbols.
  }\label{fig:figure2}
\end{figure}
We now turn our focus to the threshold-based networks introduced in
Sec.~\ref{sec:functional_networks}. Figure~\ref{fig:figure2} shows the
statistical complexity $C$ of three recurrence networks with $N=2,500$,
$N=5,000$, and $N=10,000$ nodes obtained from the R\"ossler system (Eqs.~\eqref{eqn:roesslerx}-\eqref{eqn:roesslerz}) depending on the link
density $\rho$ that is applied to obtain the recurrence matrix $\textbf R$ (Eq.~\eqref{eqn:recurrence}). For all cases, $C$ is computed as the average
statistical complexity taken over an ensemble of $n=100$ reference networks.

We note that $C$ increases with increasing $\rho$ until a maximum is reached
at $\rho_{max}=0.16$, $\rho_{max}=0.14$ and $\rho_{max}=0.13$
for $N=2,500$, $N=5,000$, and $N=10,000$, respectively (Fig.~\ref{fig:figure2}).
For values $\rho > \rho_{max}$ the statistical complexity decreases
monotonically and approaches $C=0$ for $\rho=1$ (not shown). Usually, when
constructing recurrence networks a link density of order $\mathcal{O}(10^{-2})$
is chosen heuristically, even though it was suggested that with such comparably
low choices of recurrence rates possibly significant interdependencies between
nodes might be
suppressed~\cite{donges2011identification,marwan2009complex,donner2010recurrence}.
Our results indicate that larger choices of recurrence rates and corresponding
thresholds $T$, might yield a recurrence network with higher statistical
complexity and, thus, a larger degree of non-trivial structure than the ones
that were previously typically studied. 

For the functional networks that are constructed from climate time series
across the globe according to Eqn.~\ref{eqn:climate} at different link
densities $\rho$, we find that the link densities that are maximizing the
statistical complexity are $\rho_{max}=0.15$ and $\rho_{max}=0.14$ for the
temperature and pressure field, respectively. As for the recurrence network
studied above, these values are again considerably larger then the usually
employed link densities of order
$\mathcal{O}(10^{-2})$~\cite{donges2009complex,donges_backbone_2009}.  However,
it has again been reported that these usually employed small choices of link
density and the corresponding high threshold may suppress statistically
significant signals associated with comparably lower pairwise similarity
values~\cite{wiedermann_hierarchical_2016}. Thus, future work in this area
could apply our formalism to determine a more objectively chosen threshold than
in previous studies. 

We emphasize that even though we only present two different use cases as
examples, our framework may be applicable to any kind of functional
network that is constructed from some pairwise functional interdependencies
between nodes, including neural~\cite{bullmore2009complex,bota_online_2007} or
economic networks~\cite{maluck2015network}. Beyond this, our framework might
also be applicable to networks constructed from non-pairwise interdependencies
that are investigated in, e.g, causal effect
networks~\cite{runge2015identifying,kretschmer2016using,lange_recurrence_2015}. The assessment of
statistical complexity could help to more objectively choose thresholds for the
construction of such networks and complements existing approaches based on,
e.g., the assessment of the recurrence network's percolation
threshold~\cite{eroglu2014finding,jacob2016can,donges2012analytical}.

\section{Conclusion and Outlook}\label{sec:outlook}
We have presented a methodology to categorize complex networks by means of an
entropy measure and an estimator of statistical complexity. In particular, our
method computes for each network under study an average per-node Shannon
entropy that is based on probabilities to randomly jump between neighboring
nodes in the network. From this, we estimate a network's statistical complexity
by computing the Jenson-Shannon divergence between a given network and a set of
corresponding Erd\H os-R\'enyi random networks. We find that networks of
different types, such as social or infrastructure networks, generally occupy
distinct regions in the two-dimensional complexity-entropy plane and our
proposed framework thus discriminates well between them. Moreover, we find that
connectome networks are among the statistically most complex ones while
infrastructure networks generally display a lower complexity. These properties
might intuitively be expected when considering the term complexity with respect
to real world structures and the associated functions thereof.

We have further shown in a second application that the notion of statistical
complexity can be applied to objectively estimate thresholds for
the construction of functional networks, such that a network's statistical complexity is
maximized and, hence, contains most non-trivial information. 

This work has demonstrated possible scenarios for applying the proposed
methodology. Future work should investigate in more detail the discriminating
power of the statistical complexity for a broader set of real world complex
networks. In particular, as we have observed that within our framework
connectome networks are among the most complex ones, we suggest to further
investigate the interplay between the statistical complexity and the complexity
of structure-function relations in such networks~\cite{zhou2007structure} in future
work.  Additionally, the framework should be generalized to the case of
weighted and/or directed networks. For this purpose, more emphasis must be put
into the definition of the reference networks, which for now have been assumed
to just be a randomized correspondent of the specific network under study.
Another interesting line of inquiry would be to study the dependence of the
statistical complexity with respect to the choice of the underlying random
walk, such as the maximum entropy random walk~\citep{burda_localization_2009},
as an alternative to the generic one-step random walked used in this paper. 

In general, our framework expands the understanding of complex topological
structures and helps to quantify varying degrees of complexity in various
systems. Our approach should be useful for many disciplines of
(applied) complex network science, such as neuro-, social or even climate
science.

\begin{acknowledgments}
  MW and RVD have been supported by the German Federal Ministry for Education
  and Research (BMBF) via the Young Investigators Group CoSy-CC$^2$ (grant
  no.~01LN1306A). JFD thanks the Stordalen Foundation via PB.net and the Earth
  League's EarthDoc program. MW and JFD thank the
  Leibniz Association (project DOMINOES) for financial support. JK and RVD
  acknowledge the IRTG 1740 funded by DFG and FAPESP\@. The authors thank Jobst Heitzig and
  Michael Small for fruitful discussions in the course of
  conducting this study. The European
  Regional Development Fund (ERDF), the German Federal Ministry of Education
  and Research and the Land Brandenburg have greatly supported this project by
  providing resources on the high performance computer system at the Potsdam
  Institute for Climate Impact Research.
\end{acknowledgments}

\end{document}